\documentclass[journal]{IEEEtran}
\usepackage{cite}

\ifCLASSINFOpdf
 \usepackage[pdftex]{graphicx}
\else
 \usepackage[dvips]{graphicx}
\fi
\usepackage[cmex10]{amsmath}
\usepackage{algorithm}
\usepackage[noend]{algpseudocode}

\RequirePackage[singlelinecheck=off]{caption}

\ifCLASSOPTIONcompsoc
  \usepackage[caption=false,font=normalsize,labelfont=sf,textfont=sf]{subfig}
\else
  \usepackage[caption=false,font=footnotesize]{subfig}
\fi
\usepackage{cuted}

\begin{document}
	%
	\title{Joint Audio-Video Fingerprint Media Retrieval Using Rate-Coverage Optimization}
	%
	%
	%
	
	\author{Guanghan Ning,~\IEEEmembership{Student Member,~IEEE,}
		Zhi Zhang,~\IEEEmembership{Student Member,~IEEE,}
		Xiaobo Ren,~\IEEEmembership{Member, ~IEEE,}
		Haohong Wang,~\IEEEmembership{Senior Member, ~IEEE,}
		and~Zhihai~He,~\IEEEmembership{Fellow,~IEEE}}
	\maketitle
	
	\begin{abstract}
		In this work, we propose a joint audio-video fingerprint ACR technology for media retrieval. The problem is focused on how to balance the query accuracy and the size of fingerprint, and how to allocate the bits of the fingerprint to video frames and audio frames to achieve the best query accuracy. By constructing a novel concept called Coverage, which is highly correlated to the query accuracy, we are able to form a rate-coverage model to translate the original problem into an optimization problem that can be resolved by dynamic programming. To the best of our knowledge, this is the first work that uses joint audio-video fingerprint ACR technology for media retrieval with a theoretical problem formulation. Experimental results indicate that compared to reference algorithms, the proposed method has up to 25\% query accuracy improvement while using 60\% overall bit-rates, and 25\% bit-rate reduction while achieving 85\% accuracy, and it significantly outperforms the solution with single audio or video source fingerprint. 
		
	\end{abstract}
	

	%

	\section{Introduction}
	%
	%
	%
	%
	
	\IEEEPARstart{W}{ith} widespread Internet services and the popularity of mobile devices, enormous video contents are produced and uploaded onto the Internet. For example, more than 100 hours of videos are uploaded to YouTube per minute and over 3.2 petabytes of videos have been uploaded to Vimeo \cite{chou2015pattern}. Consequently, the large amount of raw data make storing of databases in memory unfeasible.
	
	To effectively manage the rapidly growing multimedia data, a large number of methods have been proposed for multimedia content analysis and retrieval. These works include content-based image retrieval \cite{he2004manifold,he2004learning, zhang2007effective}, audio retrieval \cite{maddage2004content}, and video retrieval \cite{fan2004classview}. Given a query example provided by a user, the retrieval process is to rank the database multimedia data according to their relevance to the query example and return the top-ranked ones.
	
	For content-based multimedia retrieval \cite{lew2006content}, multimedia data are represented by feature vectors, which saves considerable storing space compared to storing raw data. However, memory is still in short supply for the large memory footprints of descriptors like SIFT \cite{ lowe2004distinctive}, SURF\cite{ bay2006surf} and GLOH\cite{mikolajczyk2005performance}. The local image descriptors computed for a reference image or object number in the thousands and each require a large number of bits to represent.
	
	In order to further compress the database and cut down the cost of database infrastructure, a number of methods have been proposed. These methods can be mainly classified into two categories： manifold learning and descriptor compression.
	
	For the first category, the methods target at exploring correlations and clustering similar features \cite{ fayyad2003multi} in order to reduce redundancy. Manifold Ranking (MR) is one of the most popular graph-based ranking methods and has been widely used for information retrieval. Due to its ability to capture the geometric structure of the image set, it has been successfully used for image retrieval. Given a query image, manifold-ranking based image retrieval （MRBIR）\cite{he2004manifold} first makes use of a manifold ranking algorithm to explore the relationship among all the data points in the feature space, and then measures relevance between the query and all the images in the database accordingly, which is different from traditional similarity metrics based on pair-wise distance. The manifold learning algorithm \cite{seung2000manifold, tenenbaum2000global, wang2013multi } has received a lot of research attention and it has been proved that the manifold structure is more powerful than Euclidean structure for data representation in many areas \cite{ he2004learning, he2004locality, he2004manifold}.
	
	For the second category, the methods \cite{duan2014compact, duan2014compact, johnson2010generalized} aim at generating compact descriptors individually, therefore reducing the overall demand for storage space. As is introduced in \cite{duan2014compact}, descriptors can be compressed by local descriptor compression \cite{ johnson2010generalized}, and global descriptor aggregation \cite{ jegou2012aggregating}. Extraction and transmission of compact descriptors are also valuable for next-generation mobile visual search applications. It can significantly reduce network latency and improve user experience. Moreover, sending compact descriptors throughout 3G network may reduce power consumption of mobile devices \cite{ girod2011mobile, ji2011towards}.
	
	The two kinds of methods are complementary and can be applied to maximize the storage reduction for multimedia databases. However, in content-based multimedia clustering area, most researches focus on data correlations within single modality, such as image clustering \cite{mclachlan1988mixture, frey2007clustering} and audio clustering \cite{ guo2003content}, but ignores cross-modal correlation between images and audios.
	
	The intrinsic problem for cross-modal correlation learning \cite{costa2014role, lu2015content, yang2008harmonizing, rasiwasia2010new, zhang2007cross} between images and audios is the obvious heterogeneity between visual features and auditory features. Images are usually represented with low-level visual features, such as color, texture, shape, etc, while audio clips are represented by different features, such as centroid, rolloff, spectral flux, and root mean square.
	In this work, we consider the design of a system that optimizes the conservation of joint video-audio descriptors. By constructing a novel concept called Coverage, which is highly correlated to the query accuracy, we are able to build a Rate-Coverage model. Given an arbitrary budget of storage space, this model aims at optimizing the retrieval accuracy while preserving only a subset of the overall joint video-audio descriptors, therefore reducing the entire storage space. We propose a dynamic programming method that resolves the optimization problem.
	
	We organize the paper's structure as follows: In section II, we briefly review the related backgrounds that are worth noting, followed by the proposed rate-coverage optimization method in section III. In section IV, we describe our implementation and illustrate our experimental results in details, which shows significant improvements over a diverse set of aspects.

	\begin{figure}[ht] 
		\centering
		\captionsetup{justification=centering}
		\includegraphics[width=3.6in]{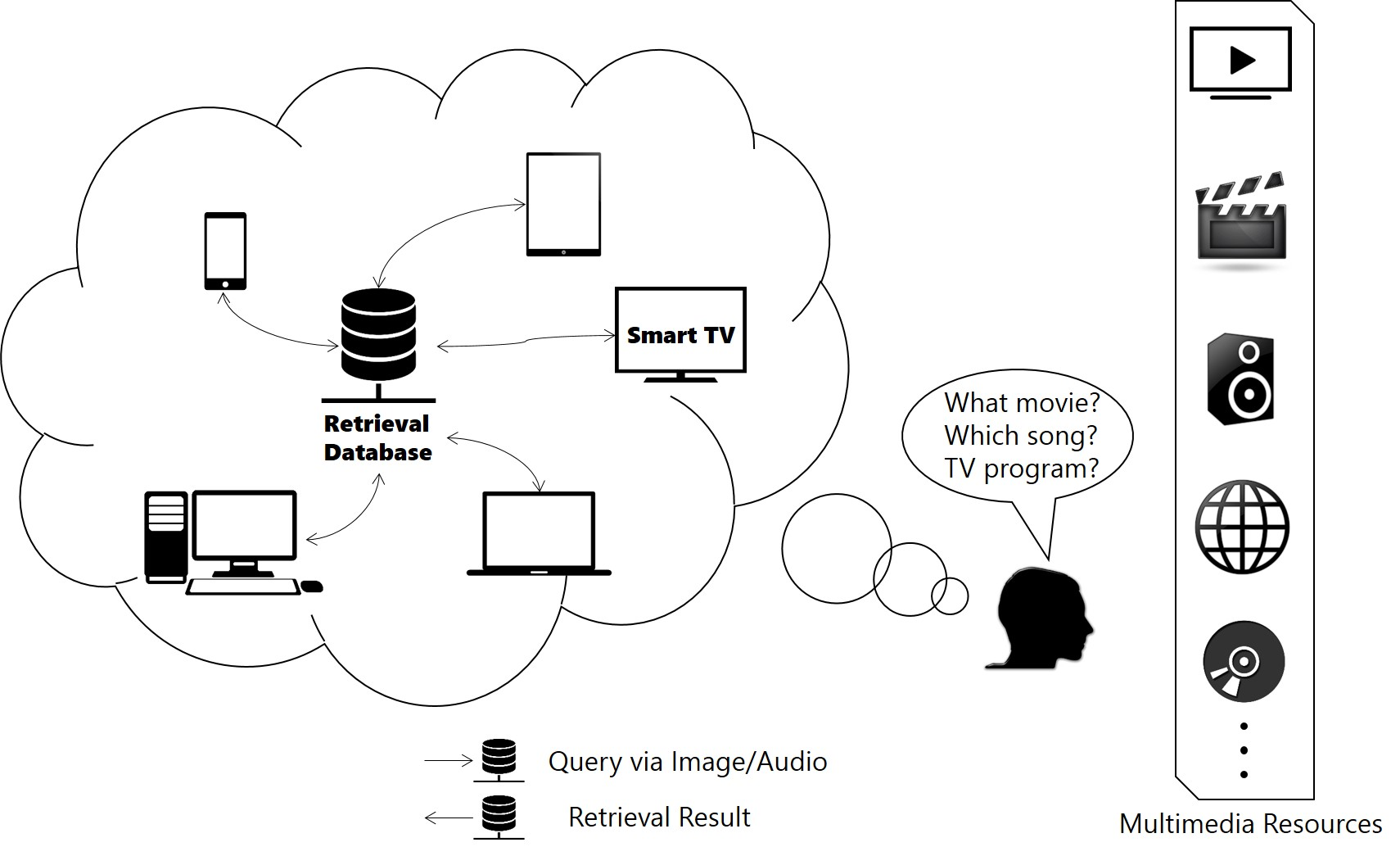}
		\caption{Typical application scenes for multimedia content based retrieval problem.}
		\label{fig-application-scene}
	\end{figure}	

	\section{Related Work} \label{related-work}
	The research community has addressed the problem of ACR by two approaches: \emph{watermarking} and \emph{fingerprinting}. Watermarking is about inserting specific identifier and time-stamp information in the content which can be viewed later. Fingerprinting, on the other hand, analyzes and compares unique content characteristic with reference database. Fingerprinting does not require the source media to be modified in any way, enables service providers to prepare indexing system for any broadcast or TV programs with no change in the content creator’s or broadcaster’s work flow. Another advantage of fingerprinting is the additive property: unlike watermarking, fingerprinting requires no copyright to the media since it does not change the original content, making it possible to be broadly used across as many applications as possible. As a result, fingerprinting system is the most widely used content recognition technique in ACR. More specifically, fingerprints, also referred to as signatures or features, are intrinsic data to characterize the video content for indexing, searching, and ranking \cite{beecks2010comparative}.  
	
	\subsection{Image fingerprinting} 
	In the last decade, Content-based Image Retrieval (CBIR) has concerned voluminous research paving way for enlargement of numerous techniques and systems besides creating interest on fields that aid these systems. In the field of content-based image retrieval \cite{datta2008image,smeulders2000content,sebe2003state}, the feature space frequently comprises position, color, or texture dimensions \cite{deselaers2008features,veltkamp2001features} where each image pixel is mapped to a single feature in the corresponding feature space. At this level, features are derived from the media without considering external semantics. A variety of low level feature methods have been developed. For example, color correlograms has been developed in CueVideo \cite{adams2002ibm}. Location, color, and texture are jointly considered in \cite{deselaers2008features,veltkamp2001content}. Low level features provide the first and enabling step for subsequent high-level feature analysis. High-level features are also called semantic features, such as timbre, rhythm, instruments, and events \cite{wei2004content}. The preferred characteristics of CBIR system are high retrieval efficiency and less computational complexity and they are the key purpose in the design of CBIR system \cite{rao2008content}.
	
	\subsection{Audio Fingerprinting} 
	Audio recognition or retrieval has been an active research topic for many years. With the availability of vast digital archives of music and commercial/non-commercial interests, there are a large amount of contributions to this field. Significant reviews can be found in the cited \cite{cano2005review,typke2005survey} and also in \cite{ozer2005perceptual}.  Regardless of different background and goals of the researches, the first step of audio retrieval invariably involves isolating a sequence of “feature” in the piece of audio, i.e. the fingerprint. The fingerprint must be sufficiently distinguishable, so that two fingerprints can be reliably torn apart or regarded similar. Along with the reliability comes the requirement for robustness against various types of distortions, such as equalizing, white noise, pitching and so on \cite{bellettini2010framework}. It tolerates users to illuminate the query as what they want, so it constructs query formulation more comprehensive and easier than key word based retrieval \cite{malik2012content}. Three of the most widely referenced approaches in this field are described here. One of the most widely used system \cite{haitsma2002highly}, uses overlapping windows of mono-audio from which to extract interesting features. Overlapping windows must be used to maintain time-shift invariance for the cases in which exact time alignment is not known. The spectral representation of the audio can be constructed in a variety of manners, Bark Frequency Cepstrum Coefficients (BFCC) are used in their study. Fingerprint is a vector of sub-fingerprints of overlapping windows in their study. An alternate approach is explored in \cite{burges2003distortion}. Their work introduces Distortion Discriminant Analysis (DDA), which is based on a variant of LDA called Oriented Principal Component Analysis (OPCA). OPCA selects a set of directions for modeling the subspace that maximizes the signal variance while minimizing the noise power. In contrast to the two schemes above, Wang \cite{wang2003industrial} proposes looking only at power-spectrogram peaks, and records the distance between them. These links constitute a very accurate identifier of each audio source, robust to distortion due to encoding or from background noise. 
	
	\subsection{Similarity measurement}
	In ACR problems, as determining similarities among data object is of crucial importance, significant body of researches have been devoted to it. Usually a set of candidate fingerprints is compared with a reference fingerprints base, identifying the matching contents by applying distance functions. Adaptive similarity measures apply a ground distance function to determine distances among feature fingerprints’ centroids in the feature space \cite{hu2008dissimilarity}. Hausdorff Distance \cite{huttenlocher1993comparing} measures the maximum nearest neighbor distance among centroids in both fingerprints, which is only based on the centroid structure of both fingerprints and does not consider any weight distribution. Hence, an extended version Perceptually Modified Hausdorff Distance was proposed \cite{park2008color} to accommodate the weight of fingerprints. Other similarity measures such as Weighted Correlation Distance \cite{hu2008dissimilarity} and Quadratic Form Distance \cite{beecks2009signature} are all applicable distance measurements. Considering flexible fingerprints forms and various recognition system configuration, distance function needs to be selected carefully according to the layouts.

	\begin{figure}[ht] 
		\centering
		\captionsetup{justification=centering}
		\includegraphics[width=3.6in]{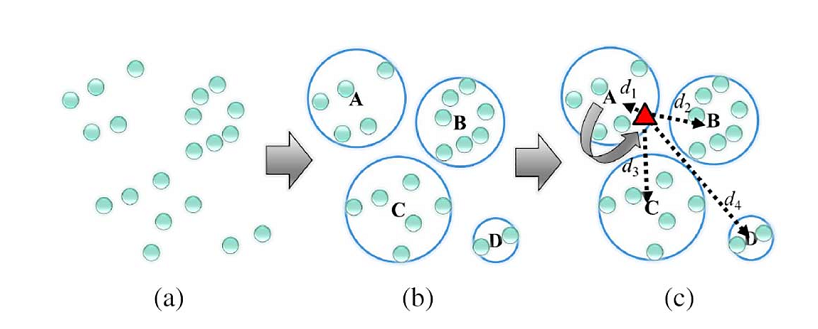}
		\caption{Example of symbol encoding. (a) Features of keyframes. (b) Reference symbol encoding. (c) Query symbol encoding according to the distances to cluster centroids.}
		\label{fig-symbol-encoding}
	\end{figure}	
	
	For near-duplicate video retrieval (NDVR) and near-duplicate video localization (NDVL), some video matching methods ultilize frame-wise visual similarity comparisons, such as dynamic programming \cite{chiu2008framework} and dynamic time warping \cite{roopalakshmi2013novel}. However, calculating similarities among all frames in a large video database is very time-consuming.  
	By considering both the spatial and temporal information, spatiotemporal methods \cite{ren2012efficient} \cite{zhang2012fast} \cite{shang2010real} \cite{chou2013near} are developed to improve the retrieval accuracy and to lower down the compurational cost. Nevertheless, their efficiency is still far from a web-scale application.
	To accelerate the frame similarity computation, as proposed in \cite{chou2015pattern}, keyframes containing similar visual features are clustered by K-means clustering, and each cluster is assigned a unique symbol. Keyframes within a cluster share the same symbol. Thereby, each reference video can be transformed into a sequence of symbols. An example of symbol encoding of reference videos is illustrated in Fig.~\ref{fig-symbol-encoding}(a) and ~\ref{fig-symbol-encoding}(b), wherein each small dot represents the feature of a keyframe in a high dimensional feature space and the large circles are the resultant clusters assigned symbols A, B, C, and D. Given a keyframe of a query video, we calculate the distance between the keyframe and each cluster centroid. The symbol(s) of the closest cluster(s) is/are assigned to the keyframe as its symbol or candidate symbols. For a keyframe of a query video (the triangle) in Fig.~\ref{fig-symbol-encoding}(c), we calculate its distance to each cluster centroid, and assign the symbol of the closest cluster, namely A in this example, to the query keyframe.

	\section{Rate Coverage Optimization} \label{rate-coverage-optimization}
	
	\begin{figure}[ht] 
		\centering
		\captionsetup{justification=centering}
		\includegraphics[width=3.6in]{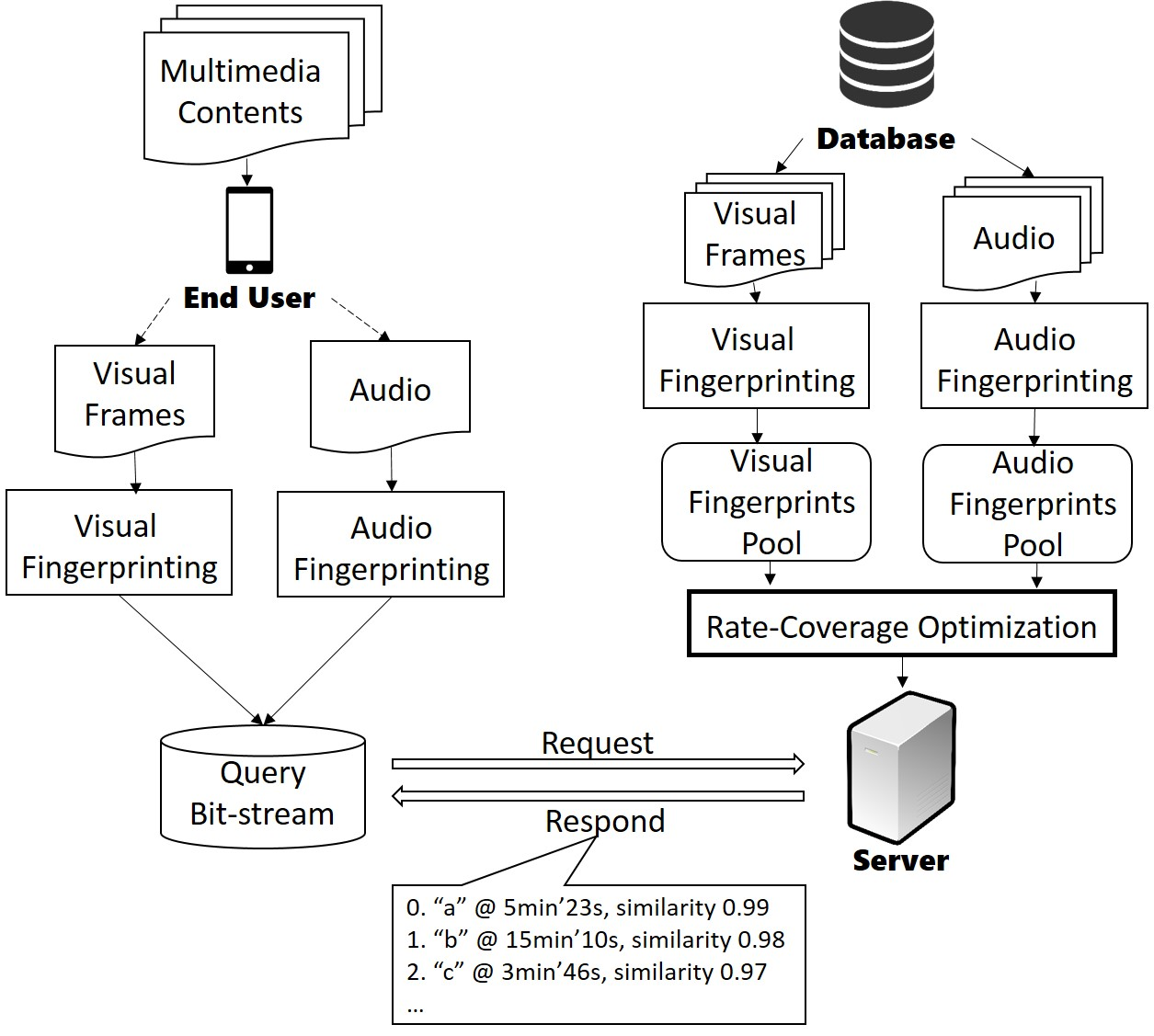}
		\caption{System Overview. Left: query route. Right: database and server with rate-coverage optimization.}
		\label{fig-sys-overview}
	\end{figure}
		
	In this work, we explore the potential of using joint audio-video fingerprints for ACR applications. In a typical usage scenario at connected home, the mobile phone can be used to detect the current TV program on the remote display by audio fingerprint-based ACR technology. However, the surrounding noises (like family chatting) could be a killer to bring the recognition accuracy down. The video fingerprint of a frame, although larger than audio fingerprint in size, is more accurate and flexible than the audio fingerprints. Nevertheless, if the ACR technology is built inside of TV by using video-based fingerprints and recognize the incoming video signals, the fingerprint database can be very large considering the 100x size of video fingerprint compared to the audio fingerprint. Therefore, it is compelling to keep the merits of audio-based and video-based ACR while overcoming their shortcomings. Thus, for an ACR system with joint representation of video and audio fingerprints, the trade-off between the fingerprint size and the retrieval accuracy in the application, have come into question. 
	
	In this paper, we aim to study in a generic ACR system, the optimized way of choosing video fingerprints over audio fingerprints, in the hope of achieving maximum expected retrieval accuracy given limited overall fingerprint size. This can be theoretically achieved considering the high redundancy residing in the correlations of successive fingerprints in feature space as well as in temporal space. We first propose a generic framework for ACR systems. As is shown in Fig.~\ref{fig-sys-overview}, the end-user device collects video and audio frames and then send the fingerprints of these frames to the query engine, which is connected to a cloud-based ACR server. On the server side, the video and audio fingerprints of large media data (some of them are online content) are generated as well using an optimization process that best balance the storage limitation and the query performance. In the following text, we focus on the optimization process, where we propose a novel way to match the problem into a rate-coverage optimization model and then solve the problem with dynamic programming technology.
	
	Without loss of generality of ACR systems, we consider the exemplary case where the user device, e.g., a smart TV, receive and decode stream from the Internet and upon user’s request, query directly with fingerprints. Note that the fingerprints are not disturbed by noise. It is also important to realize that the objective of the system is not to achieve 100\% recall, but rather to emphasize on the quality of detections, with little or no false detections, on large amounts of contents.

	\subsection{Problem Formulation} \label{Problem-formulation}
	
	In order to maximize the retrieval accuracy within storage limitation, the video and audio fingerprints must be smartly selected for the database. In this section, we transform the problem of balancing the storage limitation and the query performance to an optimization problem, which is mathematically intuitive.  
	
	The essential problem we intend to solve for the service is to maximize query accuracy $A_{t}$ with fingerprint $t$, given limited bitrate quote $R_{budget}$, where $A_{t}= 1$ if satisfactory result is retrieved, otherwise $A_{t}= 0$. The problem can then be described as:
	
	\begin{equation}   \label{eq:1}  
	Maximize \ A_{t}, t \in [0, T], \ s.t. \ R \leq R_{budget}
	\end{equation}
	
	, where $T$ is the upper limit of the possible query number.  
	
	Since the selection of incoming query from the user is a random process, we descend to evaluate the expected query accuracy considering all possible queries. The expected value of accuracy can be defined as:
	
	\begin{equation}  \label{eq:2}     
	E(A)= \dfrac{1}{T} \sum_{t=1}^{T}A_{i}
	\end{equation}
	
	Thus the original problem can be converted to the following formulation:
	
	\begin{equation} \label{eq:3}     
	Maximize \ E(A), \ s.t. \ R \leq R_{budget}
	\end{equation}                                                                                                       
	
	The data rate, denoted here by $R$, is the total amount of data we store in the database to provide the service, including both audio and video fingerprints. For video frames, the fingerprints are fixed-size feature vectors, therefore the cost of each video representative is of equal bits, denoted here by a constant $B_{V}$. Note that the audio fingerprints are natural key points of non-uniform density distribution, and that we pre-process them to derive audio segments, within which the numbers of key points are equivalent. Thus, for each audio segment, the bitrate it costs to select a representative audio frame is also uniform, denoted here by another constant $B_{A}$. The overall data rate can then be obtained by:
	
	\begin{equation}   \label{eq:4}   
	R = B_{V}\times N_{V} + B_{A} \times N_{A}
	\end{equation}  
	
	In the exemplary system, query-by-example scheme is used. Before the query process, certain amount of fingerprints are selected on the server side, from video frames and audio segments as visual and audio \textit{representatives}, respectively. The term \textit{representatives} is used to delegate the video frames and audio segments chosen to be stored in the database of the service. In equation (\ref{eq:1}), the number of video and audio representatives are denoted here by $N_{V}$ and $N_{A}$ , respectively.
	
	Each representative holds a set of K-nearest neighbors in feature space. Upon each query, we return the set with the most similar representative. In practice, we consider query result \textit{satisfactory} iff correct frame is included in the set with maximum size $K$, where $K$ is the user tolerance of how many results can be returned upon each query. In other words, if query frame is within the K-nearest neighbors of any representative frame, the query result is considered as \textit{correct}.
	
	With such definition of satisfactoriness and correctness of results, we are evaluating the query performance with a new criterion, based on our database representation. We translate the estimated \textit{Accuracy} into \textit{Coverage}, i.e., the amount of video frames and audio segments in the dataset that are \textit{correct}, or in other words, within the K-nearest neighbors of certain representatives. 
	Coverage is a property of representatives, denoted here by $C$. As will be discussed in section IV.B, this translation is justifiable as \textit{Coverage} and \textit{Accuracy} are highly correlated and affined. Therefore, equation (\ref{eq:3}) can be converted to:
	
	\begin{equation}  \label{eq:5}  
	Maximize \ C, \ s.t. \ R \leq R_{budget}
	\end{equation}  
	
	Where $C$, the fused coverage of representatives in database, is balanced between the coverage of video ($C_{V}$) and audio ($C_{A}$) via a weight control parameter $\alpha$ :
	
	\begin{equation} \label{eq:6}   
	C = \alpha C_{V,N_{V}} + (1 - \alpha)C_{A, N_{A}}
	\end{equation}  
	
	Where $\alpha \in [0, 1]$, is selected for different media sources based on experiment results, since the emphasis on video over audio, or vice versa, is not uniform over various types of content from media sources.
	
	It is important to note that the maximum coverage given the number of representatives can be found by an optimization process named Disk Covering Problem with Most $K$ points (Please see section \ref{disk-covering} for more details), which is only dependent on the number of representatives, therefore, we have:
	
	\begin{equation}  \label{eq:7}   
	C_{V,N_{V}} = f_{V}(N_{V})
	\end{equation} 
	
	\begin{equation}  \label{eq:8}   
	C_{A,N_{A}} = f_{A}(N_{A})
	\end{equation}  
	
	Where $f_{V}(N_{V})$ and $f_{A}(N_{A})$ are the optimization processes for video and audio respectively. Thus, equation (\ref{eq:5}) can be rewritten as:
	
	\begin{equation}  \label{eq:9}   
	\begin{split}
	\max_{N_{V},N_{A}}(\alpha f_{V}(N_{V}) + (1 - \alpha)f_{A}(N_{A})) \\
	subject \ to: B_{V}\times N_{V} + B_{A} \times N_{A} \leq R_{budget}
	\end{split}
	\end{equation}
	
	For various media sources, the emphasis on the importance of video and audio is diverse. The $f_{V}(N_{V})$ and $f_{A}(N_{A})$, and the rate-coverage curves are distinguishable for diverse media patterns, as will be discussed in our experiments in section \ref{exp-coverage-accuracy} and \ref{various-media-patterns}. The leverage parameter $\alpha$ plays an important role in the retrieval performance. Our proposed method achieves optimum accuracy for a given $\alpha$, but the optimum choice of $\alpha$ is inherently decided by the type of the media pattern. If the optimum choice is called $\delta$, then the choice of $\alpha$ has effects on the gap between the optimized coverage give $\alpha$ and the optimum accuracy with $\delta$. In practice, we may need to analyze the media and consider expectations of user query, in order to determine $\alpha$.

	\subsection{Disk Covering with Most K Points} \label{disk-covering}
	In order to find the maximum coverage given $N_{V}$ and $N_{A}$, we employ a RKCP3 algorithm \cite{xiao2004approximation} for optimization. The algorithm is designed to solve the \textit{Disk Partial Covering Problem}. Given $k$  disks with the same radius $r$, the partial covering issue investigates the center locations of those $k$ disks to cover the most points among total $n$ points.
	In our context, the problem is to find the $N_{V}$ number of video representatives to cover the most frames among total number of video frames, and to find the $N_{A}$ number of audio representatives to cover the most segments among total number of video segments. The radius $r$ is determined by global constraint of user tolerance to provide K-nearest neighbor, i.e., the radius is constrained by the maximum covered range for each representative, which is fixed for entire database.
	For each point $v_{i}\in V$, $G_{i}$ ($E_{i}$ respectively) is denoted as the set of points that are within distance $r$ ($3r$ , resp.) from $v_{i}$. The set $G_{i}$ is referred as representatives of radius $r$ and the set $E_{i}$ as the corresponding expanded representatives of radius $3r$. The RKCP3 algorithm procedure to cover the most video frames or audio segments can be described as follows from the original 3-approximation algorithm in \cite{xiao2004approximation}.
	
	\begin{algorithm}
		\caption{3-Approximation robust k-center algorithm}\label{pseudo-kcenter}
		\begin{algorithmic}[1]
			\Procedure{Construction of representatives}{}
			\For{$i \gets 1 \textrm{ to }k$}
			\State Let $G_i$ be the heaviest disk, i.e., contains the most uncovered points. 
			\State Mark as covered all points in the corresponding expanded disk $E_{i}$.
			\State Update all the disks and expanded disks, i.e., remove from them all covered points.
			\EndFor
			\Return $\{G_1, G_2, ..., G_k\}$
			\EndProcedure
		\end{algorithmic}
	\end{algorithm}
	
	The algorithm above has been proved to be a 3-approximation algorithm for the robust k-center problem in \cite{charikar2001algorithms}. The time complexity for the problem is $O(k\cdot n)$.

	\subsection{Optimization Solutions}
	
	The problem we intend to solve can be tackled with the following methods. However, none of them is globally optimum, except for the proposed method. In this section, we introduce the solution of these methods, with the proposed method illustrated in detail. We also explain why these methods are not optimal and why the proposed method achieves global optimum.   
	
		\subsection*{1) Arbitrary Allocation Method}
		
		The arbitrary allocation method on the allocation of bit-rates is simply ignoring the difference between audio and video, i.e., the ability to cover and the sacrifice to make, which is, the bit-rates it takes to achieve the same amount of coverage. Furthermore, it neglects the fact that video and audio usually are not of the same importance in their contribution to the coverage. As shown in Fig.  \ref{fig-rc-comparison} and Fig. \ref{fig-bitrate-allocation}, this method renders a straight line for the coverage-bitrate curve, by allocating bit-rates evenly on audio and video.         
		
		\subsection*{2) Audio First Method}
		
		The audio first method is slightly smarter than the arbitrary allocation method as it always picks audio first, based on the fact that fingerprint of a video frame is of 200 bytes, greater than the 32 bytes of an audio segment. This method renders a polyline, since the ability of the frames/audios to represent redundant ones are still ignored. 
		
		\subsection*{3) Sub-optimal Method: The Greedy Approach}
		
  		The greedy approach takes into account the representative attribute of fingerprints. Under the retrieval mechanism, representative fingerprints cover similar fingerprints in the feature space. Eqs. (\ref{eq:7}) and (\ref{eq:8}) hold true for this approach. Therefore, the maximum coverage given the number of representatives can be found by the same optimization process, i.e., \textit{Disk Covering Problem with Most K points}. Once $f_{V}(i)$ and $f_{A}(i)$ are computed with the \textit{RKCP3} algorithm, the complexity of the greedy approach is linear. As the penalty regularization parameter, $\lambda$ is also introduced. Leveraging on the weighted coverage and the corresponding bitrate cost, we obtain the leveraged gain $G$. Given limited bitrate budget, the greedy approach seeks the maximum current leveraged gain at each stage, by choosing a fingerprint type from either video or audio, until bitrate exceeds our budget.
			   
	    \begin{equation}  \label{eq:15}   
	       G = \max(\ \alpha[f_{V}(i) - f_{V}(i-1)] + \lambda B_{V}, (1 - \alpha)[f_{A}(i) - f_{A}(i-1)] + \lambda B_{A} \ )
	    \end{equation}      
			   
		The sub-optimal algorithm is more sophisticated than the methods aforesaid. We always pick the local greedy choice, that is, we choose to spare bits for a representative frame/segment by leveraging its ability to cover and the bit-rates it needs to take. Since we always choose the current best option, we are getting a fair coverage. But as we are not considering the opportunity cost of the current best, we may not be currently choosing the global best.
		
\begin{figure}[ht] 
			\centering
			\captionsetup{justification=centering}
			\includegraphics[width=2.2in]{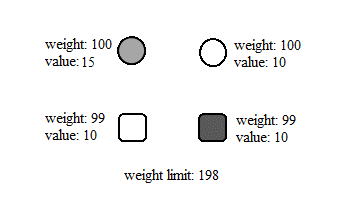}
			\caption{An example of greedy algorithm not achieving the global optimum.}
			\label{fig-greedy-example}
		\end{figure}

		Theoretically, this problem resembles the classic knapsack problem. Consider the fingerprints to be in analogy with the objects to be put in a knapsack. There are simply two kinds of objects, the video and the audio. The bit-rate of a fingerprint corresponds to the cost or weight in the knapsack problem, while its contribution to the increased coverage is analogous to the value of the object. And the limit of the overall bit-rates,  is in correspondence to the volumn of the knapsack. Consider the extreme case for the classic knapsack problem, where all of the objects are of uniform weights and of various values, the greedy approach will achieve global optimum. However, if the objects are of various weights, the greedy approach may fail to render the global optimum. One example is shown in Fig.~\ref{fig-greedy-example}, where each rectangular object is assumed to be 100 in weight and each circular object 99. The weight limit is assumed to be 198. Considering the greedy approach, we choose the circular object with value $15$, since this object is of the highest $value/weight$ ratio. However, once this object is chosen, no other object can be further selected, as is constrained by the weight limit. In contrast, the global optimum solution, is to choose the two rectangular objects, which does not exceed the weight limit but achieves a higher overall value $20$.
		
		Similarly, the greedy approach may not achieve global optimum for the proposed problem. 
		The main reason why greedy approach may not reach global optimum is, for knapsack problems, the number of objects is discrete. If the weights are continuous, the knapsack problem turns into a specific kind, called \textit{fractional knapsack problem}. In such problems, objects can be divided into pieces, each piece carrying its corresponding ratio of weight and value. For the fractional knapsack problem, the greedy approach is adequate. Apparently, for the proposed problem, the number of fingerprints is discrete as we can either store a fingerprint in the database or not.
		
		
	    \subsection*{4) Globally Optimal Method: The Proposed Approach}
	    
	    \begin{algorithm}
	    	\caption{Our proposed algorithm}\label{pseudo-global-optimum}
	    	\begin{algorithmic}[1]
	    		\Procedure{Recursive dynamic programming}{}
	    		\State Let $F_{i,j}(b)$ be the maximum coverage we can achieve given at most $i$ number of video and $j$ number of audio fingerprints, with a limitation of overall bit-rate to be $b$
	    		\State Let $I_{i,j}(b), J_{i,j}(b)$ mark the corresponding number of video and audio fingerprints that are selected as representatives while achieving coverage $F_{i,j}(y)$
	    		\For{$i \leq \overline{N_{V}}, j \leq \overline{N_{A}}$ }
	    		\State $Fs = [ F_{i-1,j}(b) , \  F_{i,j-1}(b) , \ F_{i-1,j}(b - Bv) + fv(I_{i-1,j}(b) + 1) - fv(I_{i-1,j}(b)) , \  F_{i,j-1}(b - Ba) + fa(I_{i,j-1}(b) + 1) - fa(I_{i,j-1}(b)) ]$
	    		\State $Is= [I_{i-1,j}(b), \ I_{i,j-1}(b) , \ I_{i-1,j}(b - Bv) + 1 , \ I_{i,j-1}(b-Ba) + 0]$ 
	    		\State $Js= [J_{i-1,j}(b) , \ J_{i,j-1}(b) , \ J_{i-1,j}(b - Bv) + 0 , \ J_{i,j-1}(b-Ba) + 1]$
	    		\State $index = argmax(Fs)$
	    		\State $F_{i,j}(b) = Fs(index)$
	    		\State $I_{i,j}(b) = Is(index)$
	    		\State $J_{i,j}(b) = Js(index)$
	    		\EndFor
	    		\Return $\{F_{i,j}(b), I_{i,j}(b), J_{i,j}(b)\}$
	    		\EndProcedure
	    	\end{algorithmic}
	    \end{algorithm}
	    	    
\begin{figure}[ht] 
	   		\centering
	   		\captionsetup{justification=centering}
	   		\includegraphics[width=3.6in]{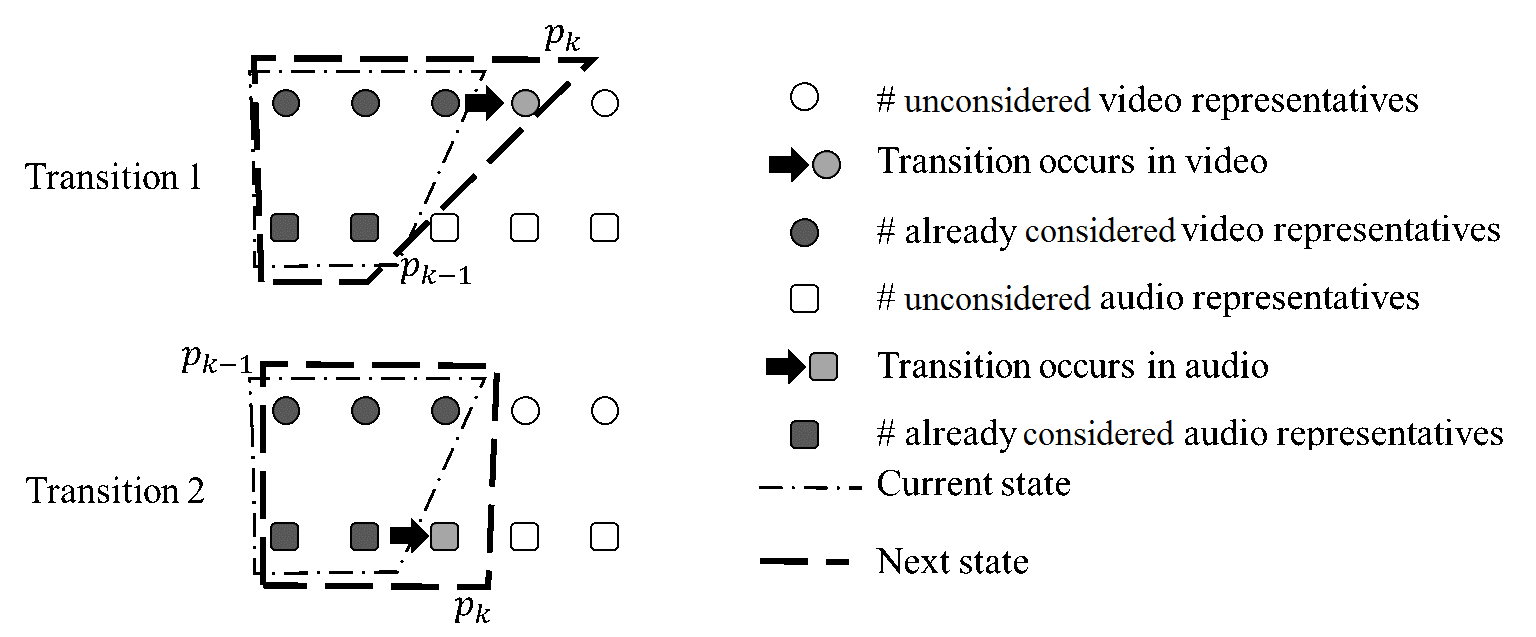}
	   		\caption{Example of transitions from previous state to current state in dynamic programming.}
	   		\label{fig-transitions}
	   	\end{figure}

	    In the following text, we explain how to solve the problem with dynamic programming, and why it achieves the global optimum.
	    We derive a solution to problem (\ref{eq:9}) using the Lagrange multiplier method to relax the bitrate constraint, so that the relaxed problem can be solved using a shortest path algorithm. We first denote the Lagrangian cost function
	    
	    \begin{equation}    \label{eq:10}  
	    J_{\lambda}(N_{V}, N_{A}) = (\alpha f_{V}(N_{V}) + (1 - \alpha)f_{A}(N_{A})) + \lambda (B_{V}\times N_{V} + B_{A} \times N_{A})
	    \end{equation}  
	    
	    where $\lambda$ is called the Lagrange multiplier. It has been proven that if there is a $\lambda^{*}$ such that
	    
	    \begin{equation}   \label{eq:11}    
	    \{N_{V}^{*}, N_{A}^{*}\} = \underset{N_{V}, N_{A}}{\operatorname{argmax}} J_{\lambda}^{*}(N_{V}, N_{A})
	    \end{equation}
	    
	    and which leads to $R = R_{budget}$, then $\{ N_{V}^{*},  N_{A}^{*} \}$ is an optimal solution to (\ref{eq:9}) . Therefore, if we can find the optimal solution to max ($J_{\lambda}(N_{V}, N_{A})$), then we can find the optimal $\lambda^{*}$ and approximation to the constrained problem of (\ref{eq:9}) .
	    As indicated in Fig.~\ref{fig-transitions}, we use a two dimensional DAG shortest Path algorithm for the optimization process, that is, in order to compute the maximal $J$, each state will need the status of $N_{V}$ and $N_{A}$ at the same time. We define a node tuple $(i, j)$ indicating state $(N_{V}, N_{A})$ in Shortest Path space, denoted as $p_{k}$, which has two paths from previous state $p_{k-1}$. It means that at this state the database stores at most $i$ number of video fingerprints and at most $j$ number of audio fingerprints as representatives. At the termination state, we derive the optimized solution for video and audio bitrate allocation, with at most $\overline{N_{V}}$ and $\overline{N_{A}}$ number of fingerprints respectively for video and audio, which is the total number of video and audio fingerprints. Therefore, starting from $(0, 0)$ to termination state $(\overline{N_{V}}, \overline{N_{A}})$, we could use dynamic programming to solve for the optimal solution. 
	    
	    To solve the optimization problem in (\ref{eq:9}) , we create a cost function $T(p_{k})$, which represents the cost to include state $(i, j)$ in the state space:
	    
	    \begin{equation}   \label{eq:12}   
	    T(p_{k}) = \max\{ \ \alpha f_{V}(i) + (1 - \alpha)f_{A}(j) + \lambda (B_{V}\times i + B_{A} \times j) \ \}           
	    \end{equation} 
	    
	    The sub-problem $f_{V}$ and $f_{A}$ are the optimization problems to maximize coverage of video and audio, given $N_{V}$ and $N_{A}$, respectively. The observation is, despite the fact that the selection of representative frames are irrelevant to the previous state, the delta cost:
	    
	    \begin{strip}
	    \begin{equation}  \label{eq:13}   
	    \Delta(p_{k-1}, p_{k}) =  \left\{
	    {\begin{array}{*{10}l}
	    	\alpha[f_{V}(i) - f_{V}(i-1)] + \lambda B_{V},  & {\begin{array}{*{10}l}
	    		if \ additional \ video \ fingerprint \\
	    		is \ selected \ during \ transition \\
	    		\end{array} }   \\
	    		    	
	    	(1 - \alpha)[f_{A}(j) - f_{A}(j-1)] + \lambda B_{A},  & \begin{array}{*{10}l}
                if \ additional \ audio \ fingerprint \\
	    	 is \ selected \ during \ transition  \end{array} \\
	    	
	    	0,  & \begin{array}{*{10}l}
	    		if \ no \ additional \ fingerprints \ is \\
	    		 selected \ during \ transition \end{array}    \\
	    		    	
	    	\end{array} } \right.
	    \end{equation}
	    \end{strip}
	    
	    , is independent of the selection of the previous states $p_{0}, p_{1}, ..., p_{k-2}$. Therefore, cost function
	    
	    \begin{equation}   \label{eq:14}   
	    T(p_{k}) = \max(T(p_{k-1}) + \Delta(p_{k-1} + p_{k}))
	    \end{equation}

	    , can be solved by a DP algorithm. In summary, the computational complexity of the DAG shortest path DP algorithm is $O(|V| + |E|)$. For the graph $G$ corresponding to the state $(i, j )$ in state space, $|V|$ is $O(N^2)$ and $|E|$ is $O(N^3)$ . Therefore, the overall computational complexity of the proposed two dimensional DAG shortest path DP algorithm is $O(N^3)O(N^3)  = O(N^6)$ where another $O(N^3)$ is the complexity of sub-problem (\ref{eq:7})-(\ref{eq:8}). Note that here we neglected the effect of computational complexity to find the optimal lambda because usually the iteration for lambda to converge is quite small compared to $O(N_{6})$ in practice.
	    In case we need to reduce the computational complexity of the optimal solution proposed above by sacrificing the system accuracy, we can use a sub-optimal solution, which is less computationally expensive by trading off with the decrease in total coverage.

	\section{Experimental Results} \label{experimental-results}
	 \subsection{Implementation Details}
	 \begin{figure*}[ht] 
	 	\centering
	 	\captionsetup{justification=centering}
	 	\includegraphics[width=6in]{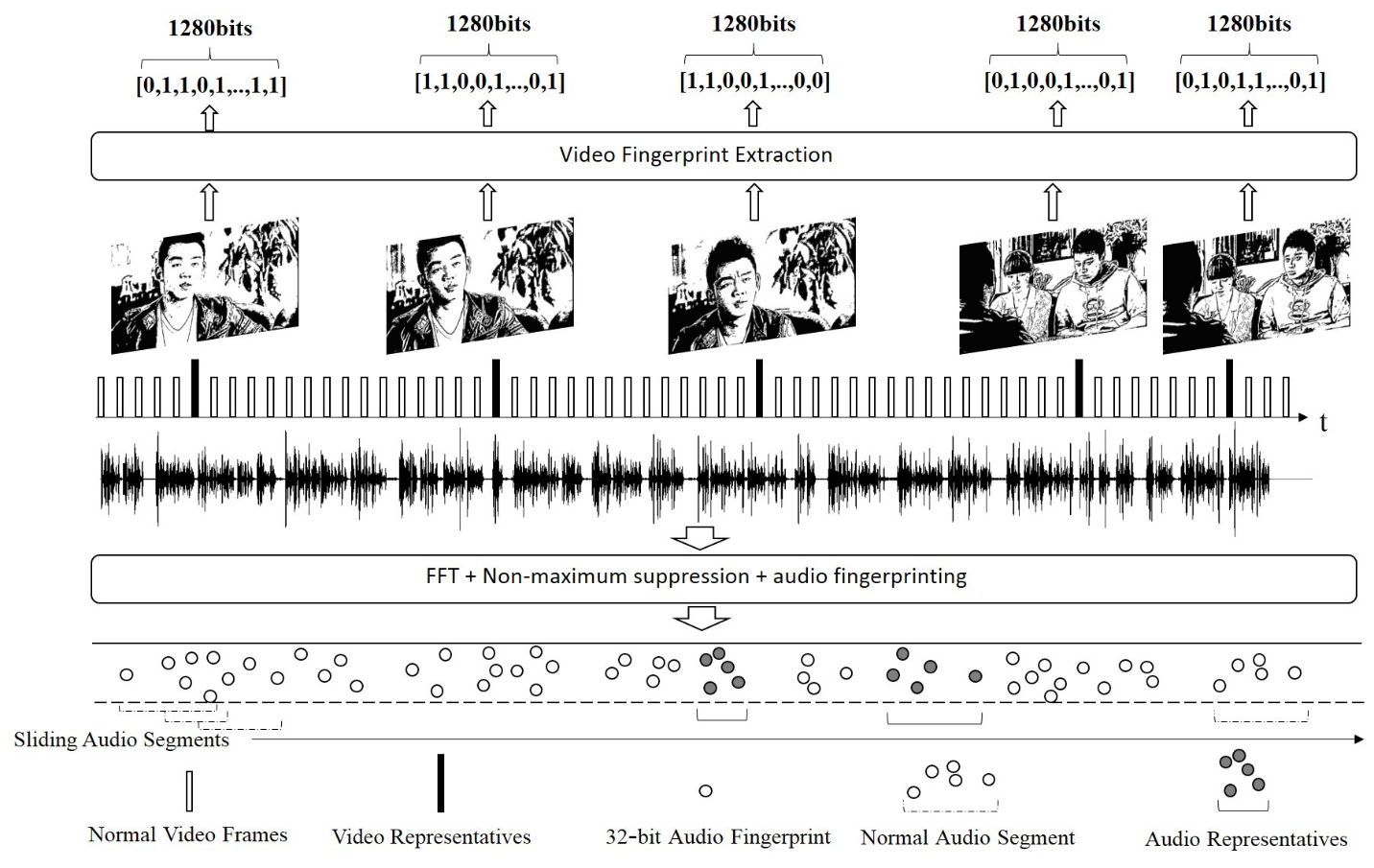}
	 	\caption{Fingerprinting process overview.}
	 	\label{fig-fingerprint-process}
	 \end{figure*}
	 
	 For the better understanding of our system, we cover some implementation details in this section. Note that this is only an example configuration, our algorithm is generic enough to accommodate any type of fingerprinting and distance function.
	 As illustrated in Fig.~\ref{fig-fingerprint-process}, the video fingerprints are extracted into groups of 1280-bits data stream while the audio fingerprints are extracted based on a key-value pair by exploring the spatial relationship among spectrogram maxims, which are computed by FFT, followed by non-maximum suppression and Shazam’s ~\cite{wang2006shazam} audio fingerprinting method. The fingerprint size of the video is therefore 160 bytes or 200 bytes depending on our experimental settings. The size of our audio segments is 32 bytes, served as unit in soundtracks.
	 
	 Similar fingerprints are mapped to the same linked list by hashing. All contents can be retrieved from database by linked list. Using different key frame as entry in list has significant impact on the indexing system and retrieval performance. 
	 
	 In our system, visual fingerprints are computed based on image pixel intensity correlations. An image frame, regardless of its aspect ratio, is first scaled to K by K square image. A selection of N out of $\bigl(\begin{smallmatrix} 2 \\N^2\end{smallmatrix}\bigr)$ block pairs are computed to generate an \emph{N}-dimensional bit array using binary comparison on pixel intensity for each image. Again an array of size M is computed based on randomly selected pixel intensities. Note that the random seed is fixed for each database during initialization, thus the fingerprinting operation is identically applied to every visual frames. The output visual fingerprints are binary strings, which are very robust to small distortion during video compression and transmission. We depict these steps in Fig.~\ref{fig-fingerprint-process}.
	 
	 \begin{figure}[ht] 
	 	\centering
	 	\captionsetup{justification=centering}
	 	\includegraphics[width=3.6in]{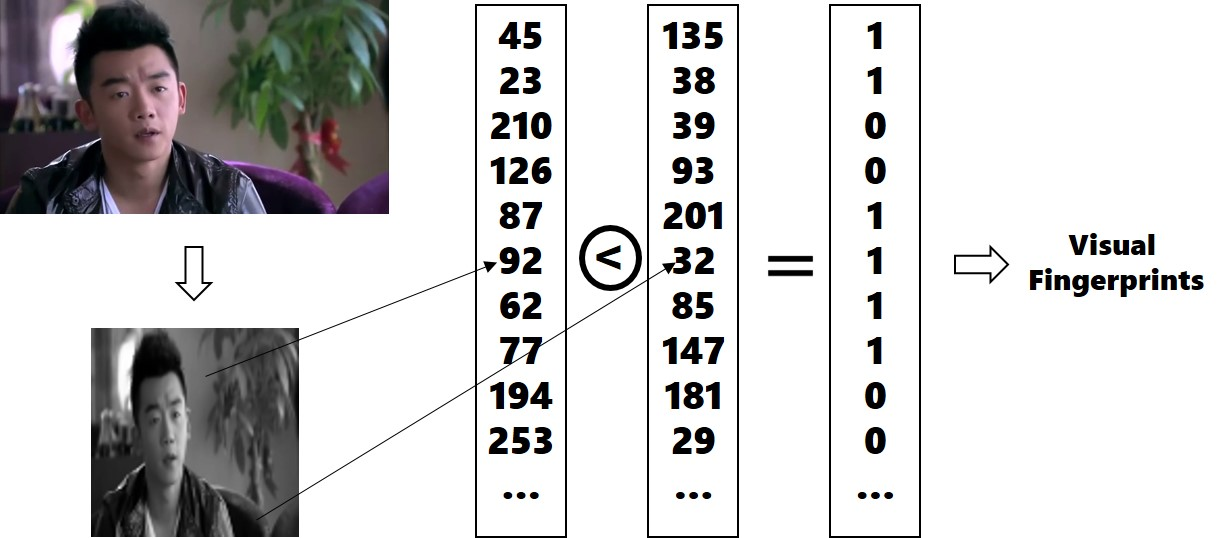}
	 	\caption{Video fingerprinting scheme.}
	 	\label{fig-video-fp}
	 \end{figure}	 
	 
	 Aside from the video content, the content of audio can be reflected from two angles. Firstly, there are measurable acoustic features from the acoustic point of view. Secondly, there are psychoacoustic features from the view of human cognition. Acoustic properties describe how sound behaves within a given physical context and they include amplitude, loudness, and etc. Psychoacoustic properties of human cognition are subjective terms perceived by the human auditory system and they include pitch, timbre, and so on. 
	 
	 Audio fingerprints, in our system are extracted on spectrograms. Each audio fingerprint unit is a combined hash key-value pair of two local maxima points on spectrogram after non-maximum suppression. For each pair, the hash key is a concatenation of $F_1$, $F_2$ and $\Delta_t$, which is the frequencies of these two points and their time domain difference, respectively. The value is the information we would like to keep upon retrieval. We use timestamp and title string as value in our system.
	 
	\begin{figure}[ht] 
		\centering
		\captionsetup{justification=centering}
		\includegraphics[width=3.6in]{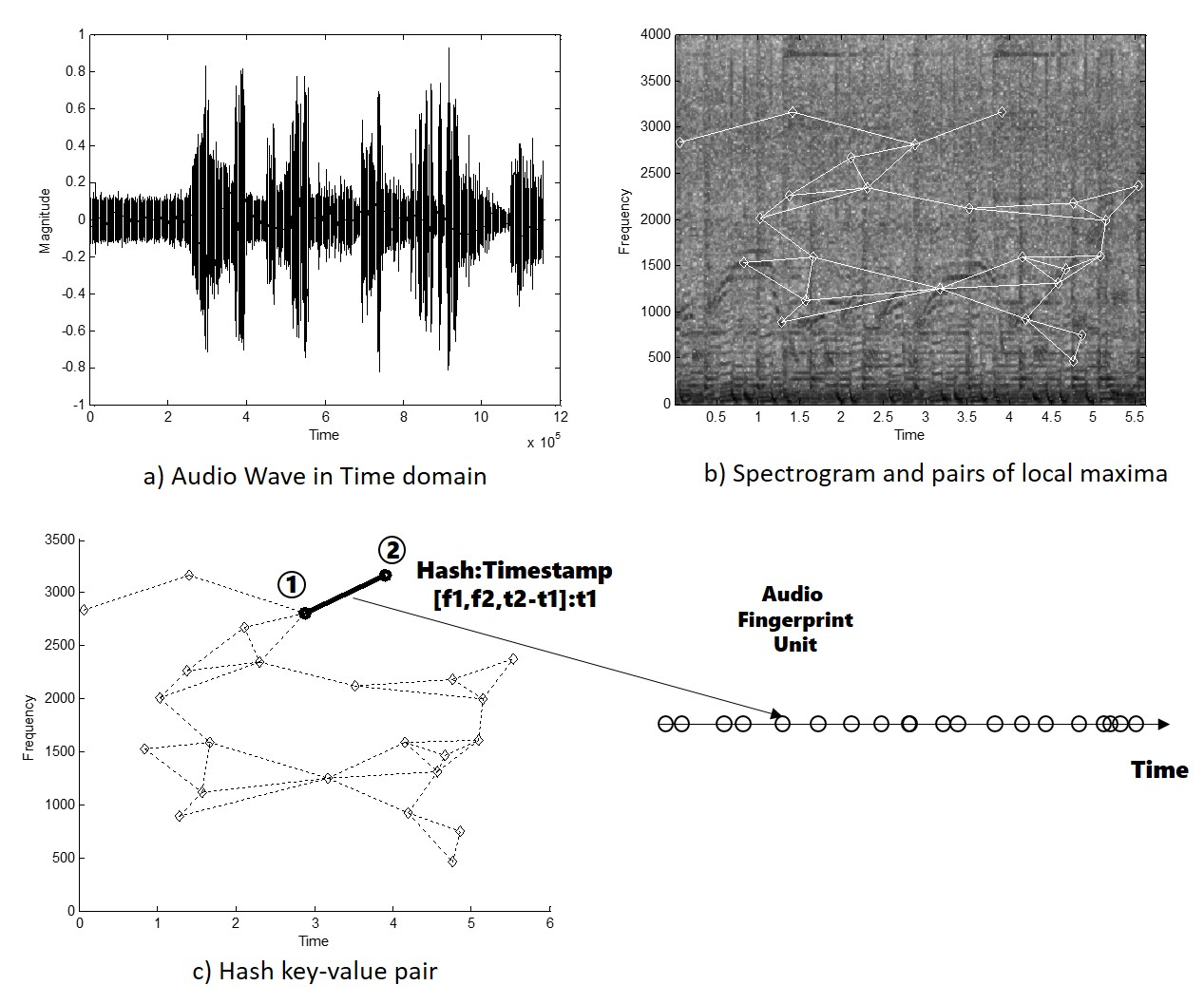}
		\caption{Audio fingerprinting sub-system. a) Original audio wave; b) Spectrograms and pairs of extracted local maxima; c) Hash-Value pairs in detail.}
		\label{fig-audio-fp}
	\end{figure}	 
	 
	 \subsection{Coverage-Accuracy Curves with fixed parameters (alpha and similarity threshold)} \label{exp-coverage-accuracy}
\begin{figure}[ht] 
		\centering
		\captionsetup{justification=centering}
		\includegraphics[width=3.6in]{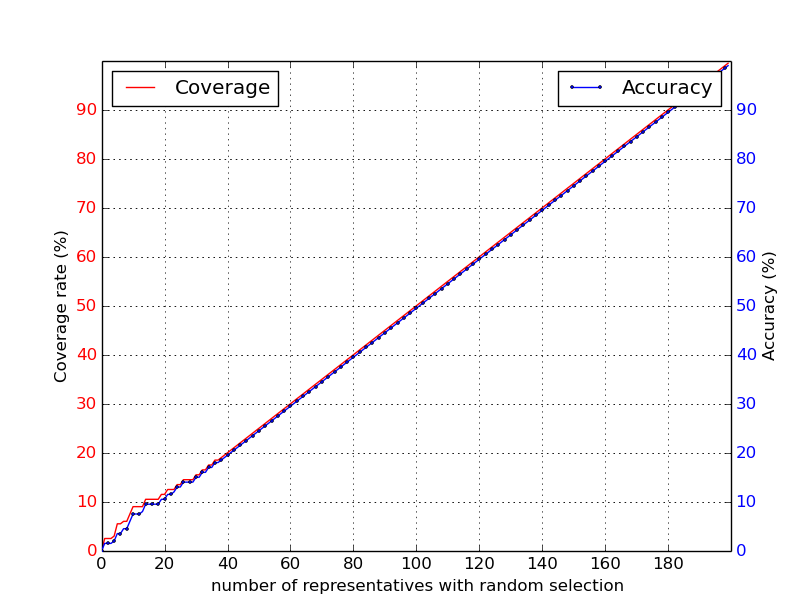}
		\caption{The coverage-accuracy relationship.}
		\label{fig-coverage-accuracy}
	\end{figure}

	 It has been discussed in \ref{Problem-formulation} that \textit{Coverage} is introduced to delegate the expected \emph{Accuracy} of the query. The accuracy is equal to the coverage, if ideally, where every query is within the top K represented frames/segments of the retrieved representative’s fingerprint cover set, rather than after the top K if there are more than K frames/segments covered by the retrieved representative. However, in practice, accuracy is slightly under the coverage level. 
	 We have conducted experiments to compute the coverage and the corresponding expected retrieval accuracy given certain bit-rate budget. By changing allocation bitrates on video and audio, we compute the coverage and the actual expected retrieval accuracy, respectively. We have conducted experiments on various media patterns. One example is shown in Fig.~\ref{fig-coverage-accuracy}, as can be observed, the coverage-accuracy curve is monotone and the gap between coverage and accuracy is very small. The coverage and accuracy is highly correlated and affined. Therefore, in order to improve retrieval accuracy, we can optimize the bit-rate allocation between audio and video to achieve optimum coverage.

	 \subsection{Various Media Patterns} \label{various-media-patterns}
\begin{figure}[ht] 
		\centering
		\captionsetup{justification=centering}
		\includegraphics[width=3.6in]{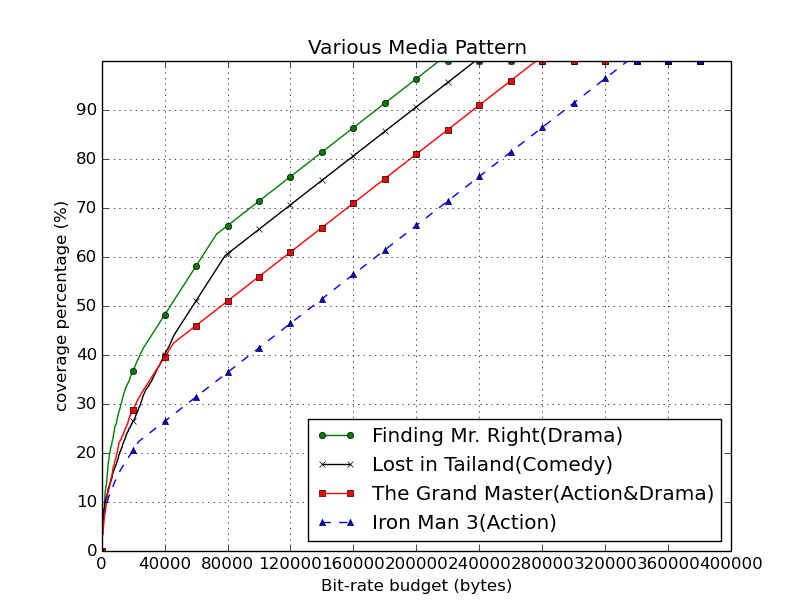}
		\caption{Rate-Coverage curves for video fingerprints only, for various media patterns.}
		\label{fig-rc-videoonly}
	\end{figure}

	 In this section, we investigate the rate-coverage model for various media patterns. We have experimented with commercial movies as various multimedia patterns and found that the movie genre does make a difference in their visual fingerprints. As one example is shown in Fig.~\ref{fig-rc-videoonly}, for various media patterns, the rate-coverage curves are distinguishable. Usually, the pace of a movie can be inferred from its genre. Slow paced movies, e.g., have longer shots on their scenes where many similar fingerprints are generated. High redundancy in fingerprints therefore results in higher coverage rate given same budget on bit-rates since redundant fingerprints can be represented by a smaller number of representatives. As we can see from Fig.~\ref{fig-rc-videoonly}, the genre makes a difference in that the fingerprints of fast-paced movies are less representational.

	 \subsection{Rate Coverage Curves for various methods}
\begin{figure}[ht] 
		\centering
		\captionsetup{justification=centering}
		\includegraphics[width=3.6in]{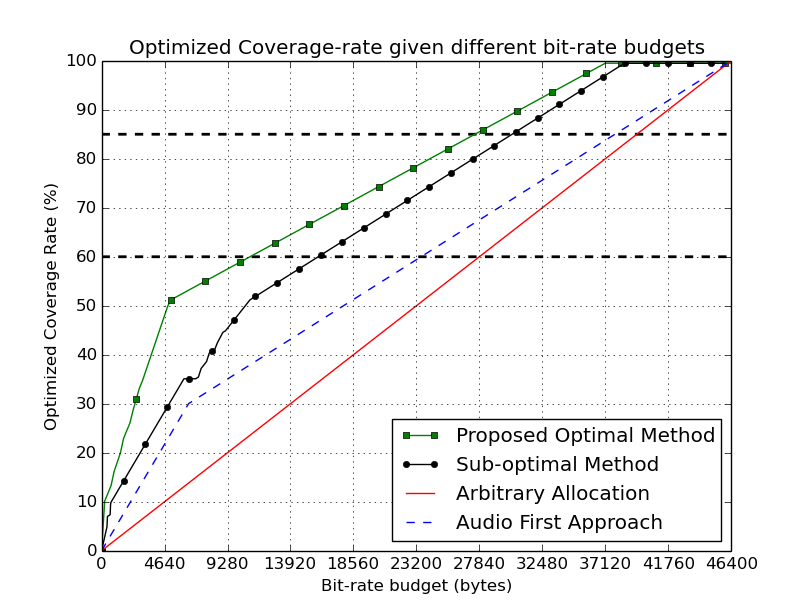}
		\caption{Rate-Coverage curves. The comparison of different approaches.}
		\label{fig-rc-comparison}
	\end{figure}

	 The effectiveness of our proposed method was tested utilizing a commercial movie, the fingerprints of which are extracted under the principles of our retrieval system. In Fig.~\ref{fig-rc-comparison}, the Rate-Coverage (RC) curves for the optimum approach and the sub-optimal approach are illustrated, together with the arbitrary allocation and audio first approaches as reference.
	 
	 Given a bitrate budget, the coverage rate of the arbitrary allocation approach is equal to the ratio of the current bit-rate budget over the total bit-rates of fingerprints. With this in mind, we can observe the performances of other approaches easily. As is shown in Fig.~\ref{fig-rc-comparison}, with very limited bit-rate budget, which is around 60\% of overall bitrate of the fingerprints, we can achieve over 85\% coverage with the proposed optimal method, with a bit-rate reduction of around 25\% compared with the arbitrary method. As a matter of fact, with only 22\% of overall bitrates as budget, we achieve 60\% of the overall coverage, saving 37.5\% bit-rate in comparison with the arbitrary method as reference.

	 As is shown in Fig.~\ref{fig-rc-comparison}, the proposed optimum algorithm outperforms other methods, which is greatly when bit-rate budget is short. With increased bit-rate budgets, the gaps between different methods are narrowed. The extreme case is, when bit-rate budget allows all the fingerprints to be stored within the database, the coverage for these methods is equally the full coverage. We also notice that with a budget of around 80\% bitrates, we can achieve full coverage with optimal approach. Therefore, there will be no need in increasing the budget further. In TABLE ~\ref{table-bitsave}, we have illustrated the average bit-rate reductions on diverse media patterns while achieving different coverages. At 95\% coverage, up to 21\% bit-rates can be saved compared to the arbitrary method.
	 
	\begin{table}[h]
		\caption[cvrg]{Average Bit-rate save of the proposed method for same coverage levels, compared with reference methods.}
		\centering
		\tiny
		\begin{tabular}{|l|l|l|l|l|l|l|}
			\hline
			cvrg                   &70\%   &75\%   &80\%   &85\%   &90\%   &95\%  \\  \hline
			Sub-optimal(Greedy)     &-9\%  &-8.5\% &-8.1\% &-8\% &-7\% &-5.5\% \\ \hline
			Audio First       &-26\%   &-25\%  &-23\% &-23\% &-21\% &-20.5\%     \\ \hline
			Arbitrary      &-32.5\%   &-31\%  &-28\%  &-26.5\% &-24\% &-21\%      \\ \hline		
		\end{tabular}
		\label{table-bitsave}
	\end{table}	 
	 
	 \subsection{Analysis of Bit-rate allocation between Video and Audio}
\begin{figure}[ht] 
		\centering
		\captionsetup{justification=centering}
		\includegraphics[width=3.6in]{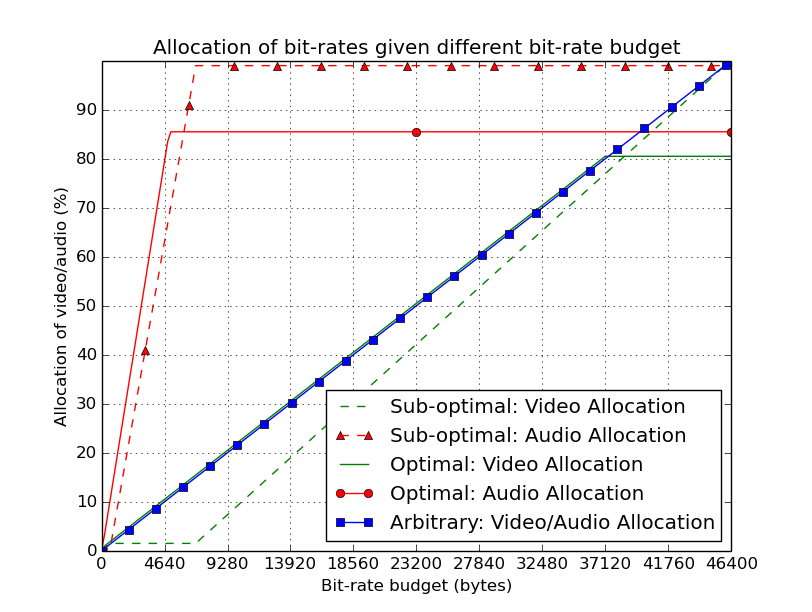}
		\caption{Bit-rate allocation of various methods over entire range of bit-rate budgets.}
		\label{fig-bitrate-allocation}
	\end{figure}

	 In this experiment, we discuss and compare how the bit-rates are allocated for different methods. The bit-rate allocation on video and audio at all the coverage levels for various methods is shown in \ref{fig-bitrate-allocation}.

	 Comparing allocations in Fig.~\ref{fig-bitrate-allocation}, we can observe the difference of allocation between these methods. The sub-optimal method inclines to allocate bytes for audio more at the beginning as it is the optimum local decision to make, because the sub-optimal approach chooses the audio as the best cost performance. Meanwhile, the increase in audio segments leads to great increase in coverage, the bit-rate cost of which is low. Comparatively, we can observe from Fig.~\ref{fig-bitrate-allocation} that the optimal approach found a better way to allocate bit-rates with optimal proportion for video and audio, which is dynamic as the budget changes. The optimal approach does not seek the current best while allocating bytes, but rather evaluates an allocation by considering the bytes left of the budget and how they can be further allocated. 
	 
	 For the optimal method, it adapts to various budgets by rectifying the proportion of bytes allocated to video and audio. In contrast, the sub-optimal method is driven by the local optimum decision. Since the video and audio is very competitive within limited budgets, the sub-optimal approach goes back and forth allocating bit-rates between video and audio. This is due to the fact that for both audio and video, with a small number of representatives we can cover dozens of other frames and segments. The sub-optimal approach chooses those representatives first, so they will run out pretty quickly, after which the increase in coverage becomes uniform with increased bit-rates.

	 \subsection{Rate-coverage curves with various Fingerprint Similarity thresholds} \label{exp-rc-curve}

	\begin{figure}[ht] 
		\centering
		\captionsetup{justification=centering}
		\includegraphics[width=3.6in]{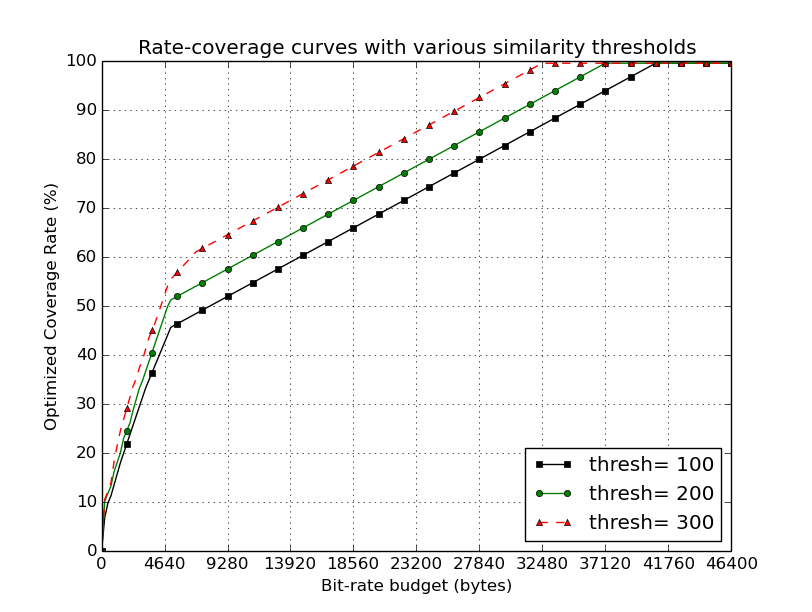}
		\caption{Rate-coverage curves with various similarity thresholds.}
		\label{fig-rc-threshold}
	\end{figure}
		 
	 In this experiment, the role of similarity threshold for fingerprints is discussed. The decision of this value should be made in accordance with our expectation for the query result. While retrieving a near-duplicate copy, the threshold is set comparatively large as the fingerprint is expected to be extremely similar. If however, semantically similar content need to be retrieved, more distortion should be tolerated. As can be observed from Fig.~\ref{fig-rc-threshold}, with higher threshold level, the optimum coverage derived from the proposed method tends to be higher. As is expected, higher threshold levels make it easier for representative fingerprints to consider other fingerprints as similar, i.e., cover them. Since representatives are more representational, the coverage increases. But note that if the threshold is too large, the gap between accuracy and coverage gets large as well, because only the K most similar fingerprints from a representative is considered successful retrieval for a query.
		 
	 \subsection{Rate-coverage curves with various leverages between video and audio}

 	\begin{figure}[ht] 
 		\centering
 		\captionsetup{justification=centering}
 		\includegraphics[width=3.6in]{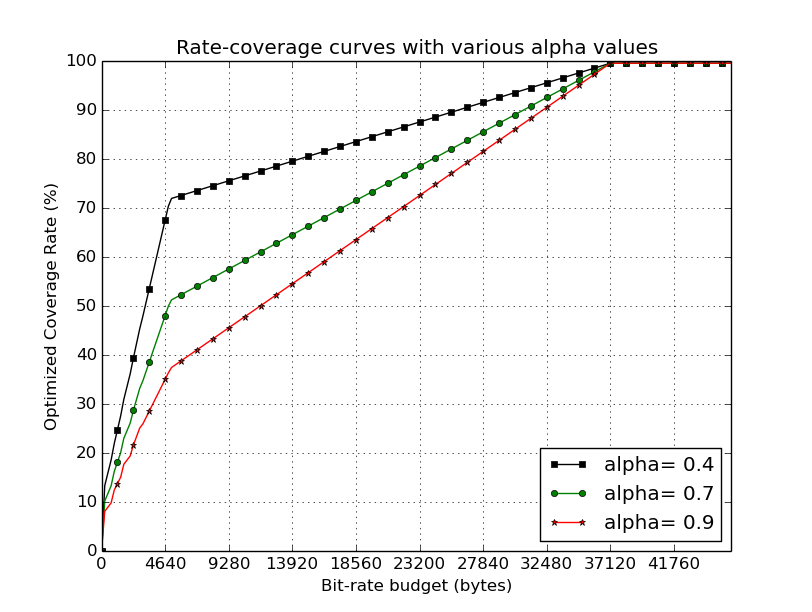}
 		\caption{Rate-coverage curves with various α values.}
 		\label{fig-rc-alpha}
 	\end{figure}
 		 
	 For various media content, the emphasis on video and audio content is diverse. Since coverage $C=\alpha C_{v,N_v} + （(1-\alpha)C_{a,N_a}$, which is positively correlated with the retrieval accuracy, it is important to analyze the effect of this value on the overall coverage. Like aforesaid, for diverse media content and patterns, their visual fingerprints are distinguishable, which influences our decision on how to leverage the importance between the two mediums.
	 
	 As illustrated in Fig.~\ref{exp-coverage-accuracy}, with fixed similarity thresholds and leverage $\alpha$ value, the retrieval accuracy is monotonically increasing with the increase of the optimized coverage, considering different bit-rate budgets. It has also been discussed in \ref{exp-rc-curve} that the rate-coverage curve is a reflection of the rate-accuracy curve, projected by the similarity threshold. In this experiment, we discuss the other parameter α, and its effect on the coverage.
	 
	 The target function of optimization is determined by $\alpha$. While emphasizing on video, the optimization process aims to allocate bit-rate to video and audio in favor of covering more video fingerprints. For media resources where video is more meaningful, e.g., a Chaplin TV show, the optimized retrieval accuracy contributes more to the user satisfaction. If however, the media source is a concert show, then more emphasis is expected to be put on audio.    
	 
	 The rate-coverage curves are plotted in Fig.~\ref{fig-rc-alpha}. These curves are generally in the same groove. As is observed, with lower emphasize on the video, we can achieve higher coverage given the same level of bit-rate budget. The more important audio is, the more bitrates are allocated for audio. Since the fingerprint of audio is less expensive than video, the number of total representatives increases, so will the total coverage be increased, as is shown in Fig.~\ref{fig-rc-alpha}.
	 
	 Note that it does not imply we are supposed to set the α value low to achieve higher retrieval accuracy, because this parameter is inherently decided by the type of the media pattern, and the coverage only reflects the accuracy in terms of this assumption of leverage.

	\section{Conclusion}
	In this paper, we proposed a novel ACR technology with joint audio-video fingerprint for media retrieval. We introduced a novel concept called Coverage to delegate retrieval accuracy, and have shown that the retrieval accuracy is positively correlated with coverage.  We utilized dynamic programming to optimize coverage for given bitrate budgets in order to maximize accuracy, and experimental results have indicated that our proposed method improves retrieval accuracy greatly compared to reference methods, and significantly saves bit-rate with same level of retrieval accuracy. Considering the difficulty of this problem, the contribution of our work is significant. In our case, fingerprints are directly from source and not disturbed by noise. The work will be completer if with noise considered. In future work, we will work on retrieving distorted fingerprints and make the proposed algorithm open to more general applications.

	
	%

	



	\newpage

	
	
	\bibliographystyle{IEEEtran}
	\bibliography{TCSVT}

\begin{thebibliography}{10}
\providecommand{\url}[1]{#1}
\csname url@samestyle\endcsname
\providecommand{\newblock}{\relax}
\providecommand{\bibinfo}[2]{#2}
\providecommand{\BIBentrySTDinterwordspacing}{\spaceskip=0pt\relax}
\providecommand{\BIBentryALTinterwordstretchfactor}{4}
\providecommand{\BIBentryALTinterwordspacing}{\spaceskip=\fontdimen2\font plus
\BIBentryALTinterwordstretchfactor\fontdimen3\font minus
  \fontdimen4\font\relax}
\providecommand{\BIBforeignlanguage}[2]{{%
\expandafter\ifx\csname l@#1\endcsname\relax
\typeout{** WARNING: IEEEtran.bst: No hyphenation pattern has been}%
\typeout{** loaded for the language `#1'. Using the pattern for}%
\typeout{** the default language instead.}%
\else
\language=\csname l@#1\endcsname
\fi
#2}}
\providecommand{\BIBdecl}{\relax}
\BIBdecl

\bibitem{chou2015pattern}
C.-L. Chou, H.-T. Chen, and S.-Y. Lee, ``Pattern-based near-duplicate video
  retrieval and localization on web-scale videos,'' \emph{Multimedia, IEEE
  Transactions on}, vol.~17, no.~3, pp. 382--395, 2015.

\bibitem{he2004manifold}
J.~He, M.~Li, H.-J. Zhang, H.~Tong, and C.~Zhang, ``Manifold-ranking based
  image retrieval,'' in \emph{Proceedings of the 12th annual ACM international
  conference on Multimedia}.\hskip 1em plus 0.5em minus 0.4em\relax ACM, 2004,
  pp. 9--16.

\bibitem{he2004learning}
X.~He, W.-Y. Ma, and H.-J. Zhang, ``Learning an image manifold for retrieval,''
  in \emph{Proceedings of the 12th annual ACM international conference on
  Multimedia}.\hskip 1em plus 0.5em minus 0.4em\relax ACM, 2004, pp. 17--23.

\bibitem{zhang2007effective}
R.~Zhang and Z.~Zhang, ``Effective image retrieval based on hidden concept
  discovery in image database,'' \emph{Image Processing, IEEE Transactions on},
  vol.~16, no.~2, pp. 562--572, 2007.

\bibitem{maddage2004content}
N.~C. Maddage, C.~Xu, M.~S. Kankanhalli, and X.~Shao, ``Content-based music
  structure analysis with applications to music semantics understanding,'' in
  \emph{Proceedings of the 12th annual ACM international conference on
  Multimedia}.\hskip 1em plus 0.5em minus 0.4em\relax ACM, 2004, pp. 112--119.

\bibitem{fan2004classview}
J.~Fan, A.~K. Elmagarmid, X.~Zhu, W.~G. Aref, and L.~Wu, ``Classview:
  hierarchical video shot classification, indexing, and accessing,''
  \emph{Multimedia, IEEE Transactions on}, vol.~6, no.~1, pp. 70--86, 2004.

\bibitem{lew2006content}
M.~S. Lew, N.~Sebe, C.~Djeraba, and R.~Jain, ``Content-based multimedia
  information retrieval: State of the art and challenges,'' \emph{ACM
  Transactions on Multimedia Computing, Communications, and Applications
  (TOMM)}, vol.~2, no.~1, pp. 1--19, 2006.

\bibitem{lowe2004distinctive}
D.~G. Lowe, ``Distinctive image features from scale-invariant keypoints,''
  \emph{International journal of computer vision}, vol.~60, no.~2, pp. 91--110,
  2004.

\bibitem{bay2006surf}
H.~Bay, T.~Tuytelaars, and L.~Van~Gool, ``Surf: Speeded up robust features,''
  in \emph{Computer vision--ECCV 2006}.\hskip 1em plus 0.5em minus 0.4em\relax
  Springer, 2006, pp. 404--417.

\bibitem{mikolajczyk2005performance}
K.~Mikolajczyk and C.~Schmid, ``A performance evaluation of local
  descriptors,'' \emph{Pattern Analysis and Machine Intelligence, IEEE
  Transactions on}, vol.~27, no.~10, pp. 1615--1630, 2005.

\bibitem{fayyad2003multi}
U.~Fayyad and J.~Shanmugasundaram, ``Multi-dimensional database record
  compression utilizing optimized cluster models,'' Oct.~14 2003, uS Patent
  6,633,882.

\bibitem{seung2000manifold}
H.~S. Seung and D.~D. Lee, ``The manifold ways of perception,'' \emph{Science},
  vol. 290, no. 5500, pp. 2268--2269, 2000.

\bibitem{tenenbaum2000global}
J.~B. Tenenbaum, V.~De~Silva, and J.~C. Langford, ``A global geometric
  framework for nonlinear dimensionality reduction,'' \emph{science}, vol. 290,
  no. 5500, pp. 2319--2323, 2000.

\bibitem{wang2013multi}
Y.~Wang, M.~A. Cheema, X.~Lin, and Q.~Zhang, ``Multi-manifold ranking: Using
  multiple features for better image retrieval,'' in \emph{Advances in
  Knowledge Discovery and Data Mining}.\hskip 1em plus 0.5em minus 0.4em\relax
  Springer, 2013, pp. 449--460.

\bibitem{he2004locality}
X.~He, D.~Cai, H.~Liu, and W.-Y. Ma, ``Locality preserving indexing for
  document representation,'' in \emph{Proceedings of the 27th annual
  international ACM SIGIR conference on Research and development in information
  retrieval}.\hskip 1em plus 0.5em minus 0.4em\relax ACM, 2004, pp. 96--103.

\bibitem{duan2014compact}
L.-Y. Duan, J.~Lin, J.~Chen, T.~Huang, and W.~Gao, ``Compact descriptors for
  visual search,'' \emph{MultiMedia, IEEE}, vol.~21, no.~3, pp. 30--40, 2014.

\bibitem{johnson2010generalized}
M.~Johnson, ``Generalized descriptor compression for storage and matching.'' in
  \emph{BMVC}, 2010, pp. 1--11.

\bibitem{jegou2012aggregating}
H.~J{\'e}gou, F.~Perronnin, M.~Douze, J.~Sanchez, P.~Perez, and C.~Schmid,
  ``Aggregating local image descriptors into compact codes,'' \emph{Pattern
  Analysis and Machine Intelligence, IEEE Transactions on}, vol.~34, no.~9, pp.
  1704--1716, 2012.

\bibitem{girod2011mobile}
B.~Girod, V.~Chandrasekhar, D.~M. Chen, N.-M. Cheung, R.~Grzeszczuk, Y.~Reznik,
  G.~Takacs, S.~S. Tsai, and R.~Vedantham, ``Mobile visual search,''
  \emph{Signal Processing Magazine, IEEE}, vol.~28, no.~4, pp. 61--76, 2011.

\bibitem{ji2011towards}
R.~Ji, L.-Y. Duan, J.~Chen, H.~Yao, Y.~Rui, S.-F. Chang, and W.~Gao, ``Towards
  low bit rate mobile visual search with multiple-channel coding,'' in
  \emph{Proceedings of the 19th ACM international conference on
  Multimedia}.\hskip 1em plus 0.5em minus 0.4em\relax ACM, 2011, pp. 573--582.

\bibitem{mclachlan1988mixture}
G.~J. McLachlan and K.~E. Basford, ``Mixture models. inference and applications
  to clustering,'' \emph{Statistics: Textbooks and Monographs, New York:
  Dekker, 1988}, vol.~1, 1988.

\bibitem{frey2007clustering}
B.~J. Frey and D.~Dueck, ``Clustering by passing messages between data
  points,'' \emph{science}, vol. 315, no. 5814, pp. 972--976, 2007.

\bibitem{guo2003content}
G.~Guo and S.~Z. Li, ``Content-based audio classification and retrieval by
  support vector machines,'' \emph{Neural Networks, IEEE Transactions on},
  vol.~14, no.~1, pp. 209--215, 2003.

\bibitem{costa2014role}
J.~Costa~Pereira, E.~Coviello, G.~Doyle, N.~Rasiwasia, G.~R. Lanckriet,
  R.~Levy, and N.~Vasconcelos, ``On the role of correlation and abstraction in
  cross-modal multimedia retrieval,'' \emph{Pattern Analysis and Machine
  Intelligence, IEEE Transactions on}, vol.~36, no.~3, pp. 521--535, 2014.

\bibitem{lu2015content}
T.~Lu, Y.~Jin, F.~Su, P.~Shivakumara, and C.~L. Tan, ``Content-oriented
  multimedia document understanding through cross-media correlation,''
  \emph{Multimedia Tools and Applications}, vol.~74, no.~18, pp. 8105--8135,
  2015.

\bibitem{yang2008harmonizing}
Y.~Yang, Y.-T. Zhuang, F.~Wu, and Y.-H. Pan, ``Harmonizing hierarchical
  manifolds for multimedia document semantics understanding and cross-media
  retrieval,'' \emph{Multimedia, IEEE Transactions on}, vol.~10, no.~3, pp.
  437--446, 2008.

\bibitem{rasiwasia2010new}
N.~Rasiwasia, J.~Costa~Pereira, E.~Coviello, G.~Doyle, G.~R. Lanckriet,
  R.~Levy, and N.~Vasconcelos, ``A new approach to cross-modal multimedia
  retrieval,'' in \emph{Proceedings of the 18th ACM international conference on
  Multimedia}.\hskip 1em plus 0.5em minus 0.4em\relax ACM, 2010, pp. 251--260.

\bibitem{zhang2007cross}
H.~Zhang, Y.~Zhuang, and F.~Wu, ``Cross-modal correlation learning for
  clustering on image-audio dataset,'' in \emph{Proceedings of the 15th
  international conference on Multimedia}.\hskip 1em plus 0.5em minus
  0.4em\relax ACM, 2007, pp. 273--276.

\bibitem{beecks2010comparative}
C.~Beecks, M.~S. Uysal, and T.~Seidl, ``A comparative study of similarity
  measures for content-based multimedia retrieval,'' in \emph{Multimedia and
  Expo (ICME), 2010 IEEE International Conference on}.\hskip 1em plus 0.5em
  minus 0.4em\relax IEEE, 2010, pp. 1552--1557.

\bibitem{datta2008image}
R.~Datta, D.~Joshi, J.~Li, and J.~Z. Wang, ``Image retrieval: Ideas,
  influences, and trends of the new age,'' \emph{ACM Computing Surveys (CSUR)},
  vol.~40, no.~2, p.~5, 2008.

\bibitem{smeulders2000content}
A.~W. Smeulders, M.~Worring, S.~Santini, A.~Gupta, and R.~Jain, ``Content-based
  image retrieval at the end of the early years,'' \emph{Pattern Analysis and
  Machine Intelligence, IEEE Transactions on}, vol.~22, no.~12, pp. 1349--1380,
  2000.

\bibitem{sebe2003state}
N.~Sebe, M.~S. Lew, X.~Zhou, T.~S. Huang, and E.~M. Bakker, ``The state of the
  art in image and video retrieval,'' in \emph{Image and Video
  Retrieval}.\hskip 1em plus 0.5em minus 0.4em\relax Springer, 2003, pp. 1--8.

\bibitem{deselaers2008features}
T.~Deselaers, D.~Keysers, and H.~Ney, ``Features for image retrieval: an
  experimental comparison,'' \emph{Information Retrieval}, vol.~11, no.~2, pp.
  77--107, 2008.

\bibitem{veltkamp2001features}
R.~C. Veltkamp, M.~Tanase, and D.~Sent, ``Features in content-based image
  retrieval systems: a survey,'' in \emph{State-of-the-art in content-based
  image and video retrieval}.\hskip 1em plus 0.5em minus 0.4em\relax Springer,
  2001, pp. 97--124.

\bibitem{adams2002ibm}
B.~Adams, G.~Iyengar, C.~Neti, H.~J. Nock, A.~Amir, H.~H. Permuter,
  S.~Srinivasan, C.~Dorai, A.~Jaimes, C.~A. Lang \emph{et~al.}, ``Ibm research
  trec 2002 video retrieval system.'' in \emph{TREC}, 2002.

\bibitem{veltkamp2001content}
R.~C. Veltkamp and M.~Tanase, ``Content-based image retrieval systems: A
  survey,'' 2001.

\bibitem{wei2004content}
C.-H. Wei and C.-T. Li, ``Content--based multimedia retrieval-introduction,
  applications, design of content-based retrieval systems, feature extraction
  and representation,'' 2004.

\bibitem{rao2008content}
C.~S. Rao and S.~Kumar, ``Content based image retrieval using contourlet
  sub-band decomposition,'' in \emph{Proceedings of IEEE International
  Conference, SPIT Colloquium, Mumbai}, 2008, pp. 140--145.

\bibitem{cano2005review}
P.~Cano, E.~Batlle, T.~Kalker, and J.~Haitsma, ``A review of audio
  fingerprinting,'' \emph{Journal of VLSI signal processing systems for signal,
  image and video technology}, vol.~41, no.~3, pp. 271--284, 2005.

\bibitem{typke2005survey}
R.~Typke, F.~Wiering, R.~C. Veltkamp \emph{et~al.}, ``A survey of music
  information retrieval systems.'' in \emph{ISMIR}, 2005, pp. 153--160.

\bibitem{ozer2005perceptual}
H.~{\"O}zer, B.~Sankur, N.~Memon, and E.~Anarim, ``Perceptual audio hashing
  functions,'' \emph{EURASIP Journal on Applied Signal Processing}, vol. 2005,
  pp. 1780--1793, 2005.

\bibitem{bellettini2010framework}
C.~Bellettini and G.~Mazzini, ``A framework for robust audio fingerprinting,''
  \emph{Journal of Communications}, vol.~5, no.~5, pp. 409--424, 2010.

\bibitem{malik2012content}
H.~Malik, ``Content-based audio indexing and retrieval: an overview,'' 2012.

\bibitem{haitsma2002highly}
J.~Haitsma and T.~Kalker, ``A highly robust audio fingerprinting system.'' in
  \emph{ISMIR}, vol. 2002, 2002, pp. 107--115.

\bibitem{burges2003distortion}
C.~J. Burges, J.~C. Platt, and S.~Jana, ``Distortion discriminant analysis for
  audio fingerprinting,'' \emph{Speech and Audio Processing, IEEE Transactions
  on}, vol.~11, no.~3, pp. 165--174, 2003.

\bibitem{wang2003industrial}
A.~Wang \emph{et~al.}, ``An industrial strength audio search algorithm.'' in
  \emph{ISMIR}, 2003, pp. 7--13.

\bibitem{hu2008dissimilarity}
R.~Hu, S.~R{\"u}ger, D.~Song, H.~Liu, and Z.~Huang, ``Dissimilarity measures
  for content-based image retrieval,'' in \emph{Multimedia and Expo, 2008 IEEE
  International Conference on}.\hskip 1em plus 0.5em minus 0.4em\relax IEEE,
  2008, pp. 1365--1368.

\bibitem{huttenlocher1993comparing}
D.~P. Huttenlocher, G.~Klanderman, W.~J. Rucklidge \emph{et~al.}, ``Comparing
  images using the hausdorff distance,'' \emph{Pattern Analysis and Machine
  Intelligence, IEEE Transactions on}, vol.~15, no.~9, pp. 850--863, 1993.

\bibitem{park2008color}
B.~G. Park, K.~M. Lee, and S.~U. Lee, ``Color-based image retrieval using
  perceptually modified hausdorff distance,'' \emph{Journal on Image and Video
  Processing}, vol. 2008, p.~4, 2008.

\bibitem{beecks2009signature}
C.~Beecks, M.~S. Uysal, and T.~Seidl, ``Signature quadratic form distances for
  content-based similarity,'' in \emph{Proceedings of the 17th ACM
  international conference on Multimedia}.\hskip 1em plus 0.5em minus
  0.4em\relax ACM, 2009, pp. 697--700.

\bibitem{chiu2008framework}
C.-Y. Chiu, C.-S. Chen, and L.-F. Chien, ``A framework for handling
  spatiotemporal variations in video copy detection,'' \emph{Circuits and
  Systems for Video Technology, IEEE Transactions on}, vol.~18, no.~3, pp.
  412--417, 2008.

\bibitem{roopalakshmi2013novel}
R.~Roopalakshmi and G.~R.~M. Reddy, ``A novel spatio-temporal registration
  framework for video copy localization based on multimodal features,''
  \emph{Signal processing}, vol.~93, no.~8, pp. 2339--2351, 2013.

\bibitem{ren2012efficient}
J.~Ren, F.~Chang, T.~Wood, and J.~R. Zhang, ``Efficient video copy detection
  via aligning video signature time series,'' in \emph{Proceedings of the 2nd
  ACM International Conference on Multimedia Retrieval}.\hskip 1em plus 0.5em
  minus 0.4em\relax ACM, 2012, p.~14.

\bibitem{zhang2012fast}
J.~R. Zhang, J.~Y. Ren, F.~Chang, T.~L. Wood, and J.~R. Kender, ``Fast
  near-duplicate video retrieval via motion time series matching,'' in
  \emph{Multimedia and Expo (ICME), 2012 IEEE International Conference
  on}.\hskip 1em plus 0.5em minus 0.4em\relax IEEE, 2012, pp. 842--847.

\bibitem{shang2010real}
L.~Shang, L.~Yang, F.~Wang, K.-P. Chan, and X.-S. Hua, ``Real-time large scale
  near-duplicate web video retrieval,'' in \emph{Proceedings of the
  international conference on Multimedia}.\hskip 1em plus 0.5em minus
  0.4em\relax ACM, 2010, pp. 531--540.

\bibitem{chou2013near}
C.-L. Chou, H.-T. Chen, Y.-C. Chen, C.-P. Ho, and S.-Y. Lee, ``Near-duplicate
  video retrieval and localization using pattern set based dynamic
  programming,'' in \emph{Multimedia and Expo (ICME), 2013 IEEE International
  Conference on}.\hskip 1em plus 0.5em minus 0.4em\relax IEEE, 2013, pp. 1--6.

\bibitem{xiao2004approximation}
B.~Xiao, J.~Cao, Q.~Zhuge, Y.~He, and E.~H. Sha, ``Approximation algorithms
  design for disk partial covering problem,'' in \emph{Parallel Architectures,
  Algorithms and Networks, 2004. Proceedings. 7th International Symposium
  on}.\hskip 1em plus 0.5em minus 0.4em\relax IEEE, 2004, pp. 104--109.

\bibitem{charikar2001algorithms}
M.~Charikar, S.~Khuller, D.~M. Mount, and G.~Narasimhan, ``Algorithms for
  facility location problems with outliers,'' in \emph{Proceedings of the
  twelfth annual ACM-SIAM symposium on Discrete algorithms}.\hskip 1em plus
  0.5em minus 0.4em\relax Society for Industrial and Applied Mathematics, 2001,
  pp. 642--651.

\bibitem{wang2006shazam}
A.~Wang, ``The shazam music recognition service,'' \emph{Communications of the
  ACM}, vol.~49, no.~8, pp. 44--48, 2006.

\end{thebibliography}
\end{document}